\documentclass[doublecol]{epl2} 

\usepackage{graphicx}
\usepackage{bm}
\usepackage{amsfonts}
\usepackage{amsmath}
\usepackage[colorlinks=true,citecolor=blue]{hyperref}
\hypersetup{colorlinks=true,citecolor=blue,linkcolor=red,urlcolor=blue}

\title{Inhomogeneous superconductivity in the presence of time-reversal symmetry}

\author{Mir Vahid Hosseini}
\shortauthor{Mir Vahid Hosseini}

\institute{Department of Physics, Faculty of Science, University of Zanjan, Zanjan 45371-38791, Iran}

\pacs{74.20.-z}{Theories and models of superconducting state}
\pacs{74.81.-g}{Inhomogeneous superconductors and superconducting systems, including electronic inhomogeneities}
\pacs{73.22.Pr}{Electronic structure of graphene}

\abstract{
We propose a new type of non-uniform condensate state in the presence of time-reversal symmetry. The underlying platform of this state is a corrugated honeycomb lattice which exhibits an inhomogeneous pseudo magnetic field. Using self-consistent tight binding Bogoliubov-de Gennes formalism, we show that an s-wave pairing of chiral carriers in the presence of an inhomogeneous pseudo magnetic field results in a spatially modulated order parameter with the half wavelength of the corrugation. The manner of modulation depends on the fundamental directions of honeycomb lattice, the zigzag and armchair directions. The stability of this inhomogeneous superconductivity is also analyzed and possible experimental realization is discussed.}

\begin{document}

\maketitle

{\it Introduction.}---
Inhomogeneous superconductivity (IS) with a spatially modulated order parameter has attracted a great deal of attention from both the theoretical~\cite{FFLOTheo} and experimental~\cite{FFLOExp1,FFLOExp2,FFLOExp3,FFLOExp4} points of view. The IS can emerge through a non-trivial pairing between electronic states. The first type of IS was considered by Fulde and Ferrell~\cite{FF}, and Larkin and Ovchinnikov~\cite{LO} (FFLO). In the FFLO state, the Cooper pairs can be formed through coupling between the imbalanced opposite spin states near the Fermi surface with non zero center-of-mass momentum, resulting in an oscillatory behavior of pairing potential in the real space. The other types of IS are gapped~\cite{TopoGappedFFLO1,TopoGappedFFLO2,TopoGappedFFLO3,TopoGappedFFLO4} and gapless~\cite{TopoGaplessFFLO1,TopoGaplessFFLO2} topological FFLO states which are protected against disorders. In contrast to the above mentioned ISs which are made by the magnetic ingredient, the possibility of IS in the absence of magnetic field is indeed an intriguing issue. So the interesting question arises, how is it possible to obtain IS in the presence of time-reversal symmetry?
\begin{figure}[t]
  \includegraphics[width=8cm]{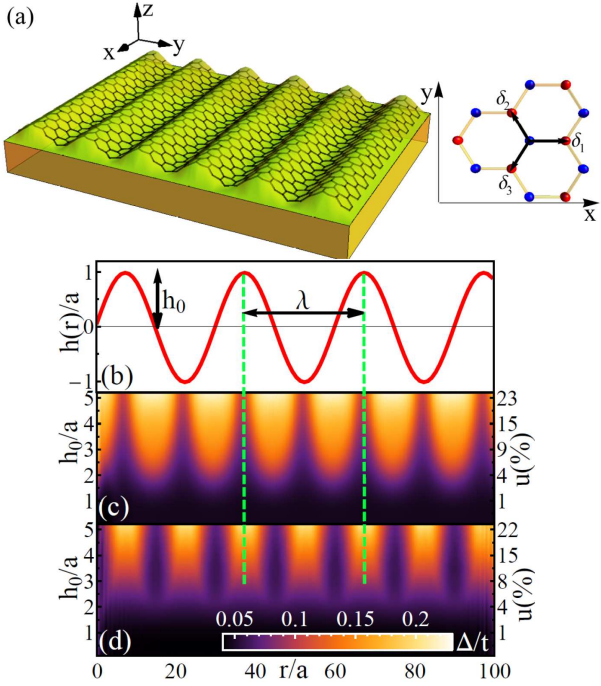}
  \caption{(Color online) (a) Left: 1D rippled graphene's surface. Right: three nearest-neighbor vectors $\mathbf{\delta}_{1}$, $\mathbf{\delta}_{2}$ and $\mathbf{\delta}_{3}$, with the geometry chosen so that the x-axis is along the armchair direction. (b) Side view of out-of-plane deformation of graphene surface with amplitude $h_0$ and wavelength $\lambda$. Spatial profiles of modulating order parameter of (c) armchair-corrugated graphene (ACG) and (d) zigzag-corrugated graphene (ZCG) in terms of ripple's amplitude $h_0$ and strain $u$. Here $g_0=1.36t$ and $\mu=0.6t$.}
  \label{cor}
\end{figure}
\par
The honeycomb lattice structure with its exceptional electronic properties has recently become an interesting subject of studies due to the diversity of natural materials such as graphene~\cite{graphene1,graphene2} and artificial structures~\cite{artificial} which can be expressed by triangular symmetry. In addition to the existence of Dirac-like energy-momentum dispersion relation that emerges from the band structure of a honeycomb lattice and resembles chiral carriers of high energy physics, the other amazing feature is that an effective gauge field can be introduced by band engineering through honeycomb lattice deformation, resulting in quantum Hall effect~\cite{grapheneHall1,grapheneHall2,grapheneHall3} in the absence of a real magnetic field. Unlike the real magnetic field, the pseudo magnetic field generated by lattice distortions preserves time-reversal symmetry. The interplay between pseudo magnetic field and superconductivity may lead to IS which has not been yet explored and we will discuss below. Remarkably, the proposed IS here is not restricted to the solid state physics and it may be supported in various systems, including color superconductivity in dense quark matter and also ultracold atoms in optical lattices. So, without loss of generality, we employ graphene as a theoretical simulator.
\par
In the present Letter, we consider a corrugated monolayer graphene, as depicted schematically in Fig.~\ref{cor}(a) (left panel), in the presence of an intrinsic s-wave superconductivity. The corrugation of graphene surface gives rise to the change of the hopping energy, leading to the presence of a non-uniform pseudo magnetic field. While most of the studies on superconducting state in strained graphene have been focused on uniform pseudo magnetic field~\cite{UniStrain1,UniStrain2,UniStrain3,UniStrainUchoa}, here, we address the question of superconducting state in the presence of a non-uniform pseudo magnetic field. We find that the ripple-induced pseudo magnetic field, interestingly, results in a spatially modulated superconducting state with the half wavelength of the corrugation, as one can see in Figs.~\ref{cor}(c) and ~\ref{cor}(d). This IS can be described with the aid of a tight-binding model in which ripples shift the energy levels of band structure towards each others, such that the density of states and degeneracy of states increase. The shape of modulation of the superconducting order parameter depends on the zigzag or armchair directions of the corrugation. Also, we find that the superconducting gap increases with increasing the corrugation amplitude. With the further increasing of corrugation amplitude, the superconducting state become more stable, suggesting that engineering a certain inhomogeneous strain to correspond to a certain pseudo magnetic field gives rise to the IS with a specific pattern.

{\it Model.}---In our model, we consider one-dimensional (1D) periodic ripples along the fundamental directions of graphene {\it i.e.}, zigzag and armchair directions, in the presence of s-wave superconducting correlations. The case of zigzag-corrugated graphene (ZCG) is depicted schematically in Fig. ~\ref{cor}(a)(left panel). The $\pi$ electrons of this system can be described by the following lattice Hamiltonian~\cite{grapheneSuperTher1,intrinsicSupe2,grapheneSuperTher2},

\begin{eqnarray}\label{HT}
H\!\!&=&\!\!H_{kin}+H_{\Delta},\\
H_{kin}\!\!&=&\!\!-\mu\sum_{i,\sigma}n_{i\sigma}-\!\sum_{i,\sigma}\!\sum^3_{\alpha=1}(t_{i\alpha}a_{i\sigma}^{\dagger}
b_{i+\mathbf\delta^{\prime}_{i\alpha}\sigma}+\textrm{H.c.}),\nonumber\\
H_{\Delta}\!\!&=&\!\!\sum_{i}\!\left[\Delta_{i}(a_{i\uparrow}^{\dagger}a_{i\downarrow}^{\dagger}+b_{i\uparrow}^{\dagger}b_{i\downarrow}^{\dagger})
+\textrm{H.c.}\right],\nonumber
\end{eqnarray}
where $a_{i,\sigma}(b^{\dagger}_{i,\sigma})$ denotes on-site annihilation (creation) operator for electron of sublattice $A$ ($B$) with spin $\sigma=\uparrow,\downarrow$ at unit cell $i$. The kinetic Hamiltonian $H_{kin}$ includes the on-site particle density operator $n_{i\sigma}$, the graphene chemical potential $\mu$ and the modified hopping energy $t_{i\alpha} = t \exp[-\beta (|\mathbf{\delta}^{\prime}_{i\alpha}|/a_{0}-1)]$~\cite{Hopp0} due to lattice deformation, where $t\approx 2.8$ ev is the equilibrium value of hopping energy, $a_{0}$ = 0.142 nm being the equilibrium value of bond length, $\beta = 3.37$~\cite{Hopp1,Hopp2} is the changes of hopping energy relative to the changes of the bond length and $\mathbf{\delta}^{\prime}_{i\alpha}$'s are three position dependent deformed nearest-neighbor vectors which their corresponding undeformed vectors are $\mathbf{\delta}_{1}=a_{0}(1,0)$, $\mathbf{\delta}_{2} = a_{0}(-1/2,\sqrt{3}/2)$ and $\mathbf{\delta}_{2} = a_{0}(-1/2,-\sqrt{3}/2)$ (see the right panel of Fig.~\ref{cor}(a)).
$H_{\Delta}$ is the on-site spin-singlet s-wave attractive interaction between electrons in graphene layer with the gap function $\Delta_{i} = -g_{0}(\langle a_{i\downarrow}a_{i\uparrow}\rangle+\langle b_{i\downarrow}b_{i\uparrow}\rangle)/2$, where $g_{0}$ is the intrinsic pairing strength. The possible mechanisms of intrinsic superconducting
correlations in single and multilayer graphene have been discussed in Ref\cite{intrinsicSupe1,intrinsicSupe2,intrinsicSupe3,intrinsicSupe4,intrinsicSupe5,intrinsicSupe6,intrinsicSupe7,intrinsicSupe8}.
We ignore the in-plane displacement of lattice points and assume that the out-of-plane deformation of the graphene layer can be approximated by a sinusoidal function $h(r) = h_0 \sin(2\pi r/\lambda)$~\cite{sin1,sin2,sin3}, where $h_0$ is the ripple height, $\lambda$ is its wavelength and the in-plane lattice position $r$ is equivalent to $x$ and $y$ for the case of armchair-corrugated graphene (ACG) and ZCG, respectively. It should be noted that the main change of hopping comes through the strain and the effects of curvature on hopping energies contribute negligibly, so the formula for the hopping only involves the bond length modulation \cite{graphene2}. We define strain $u$ as the elongation of graphene in the direction of the corrugation per relaxed one due to out-of-plane deformation. Note that the superconducting order parameter is taken to be the average of the pair amplitude of sublattices A and B in each unit cell, {\it i.e.}, $\langle a_{i\downarrow}a_{i\uparrow}\rangle$ and $\langle b_{i\downarrow}b_{i\uparrow}\rangle$, this assumption is valid in the case $h_0 \ll \lambda$ which is correspond to small strain limits.
\par
The $x$-axis is taken to be in the armchair direction (see right panel of Fig.~\ref{cor}(a)). We apply open boundary conditions (OBCs) along corrugation direction and assume that translational invariance is preserved in a direction perpendicular to the corrugation direction. This implies that for the ACG (ZCG), crystal momentum $k_y$ ($k_x$) is a good quantum number. Adopting periodic boundary conditions (PBCs) for translational invariant direction, we can perform the Fourier transformation from Eq. (\ref{HT}). Using the Bogoliubov-Valatin transformation~\cite{Bogoliubov-Valatin1,Bogoliubov-Valatin2,Bogoliubov-Valatin3}, the Hamiltonian can be diagonalized by solving the following tight binding Bogoliubov-de Gennes equations self-consistently,
\begin{align}\label{eq:diag}
\!\!\sum_m\!\!&\begin{pmatrix}
\!\mathcal{H}^{ACG(ZCG)}_{m,n}\! & \!\Delta_{m,n}\!\! \\
\!\Delta^\dag_{m,n}\! &\! -\mathcal{H}^{ACG(ZCG)}_{m,n}\!\! \\
\end{pmatrix} \!\psi^\nu_m\!\!=\!\!E^\nu(k_{y(x)})\psi^\nu_n,
\end{align}
where $\Delta_{m,n}\!\!=\!\!\Delta_{n}\delta_{m,n}\sigma_0$, $n$ indexes the lattice site along the corrugation, $\delta_{m,n}$ is the Kronecker delta function and $\sigma_0$ is the identity matrix. The crystal momentums $k_x =2\pi l/(\sqrt{3}N_x\textrm{a})$ and $k_y = 2\pi l/(N_y\textrm{a})$ with integer $l$ are chosen such that their values remain in the first Brillouin zone. $N_x$ and $N_y$ denote the number of unit cells and $\textrm{a} = \sqrt{3}a_0$ is the lattice constant. Also, $E^\nu(k_{y(x)})$ and
$\psi^\nu_n = [u_n^\nu(k_{y(x)}), y_n^\nu(k_{y(x)}), v_n^\nu(k_{y(x)}), z_n^\nu(k_{y(x)})]^\text{T}$ are eigenvalues and eigenfunctions, respectively. We have also defined the normal matrices
$\mathcal{H}^{\!ACG}_{\!m,n}\!\!\!=\!\!\!-t_{m,1}\delta_{m,n}\sigma_x
\!\!-\!\!\sum_{\xi=\pm}\!\!\delta_{m+\xi,n}[t_{m,2}\exp(\frac{i\xi k_y\textrm{a}}{2})\!\!+\!t_{m,3}\exp(\frac{-i\xi k_y\textrm{a}}{2})]\sigma_{\!\xi} $
for the case of ACG and
$\mathcal{H}^{\!ZCG}_{\!m,n}\!\!\!\!=\!\!\!-\!\!\sum_{\xi=\pm}\![t_{m,1}\delta_{m,n}\exp(\frac{i\xi k_x\textrm{a}}{2\sqrt{3}})\!\!+\!\!(t_{m,2}\delta_{m+\xi,n}\!+\!t_{m,3}\delta_{m-\xi,n})\exp(\frac{-i\xi k_x\textrm{a}}{2\sqrt{3}})]\sigma_{\!\xi} $ for the case of ZCG. Here $\sigma_{\pm}=(\sigma_x\!\pm\!i\sigma_y)/2$, $\sigma_x$ and $\sigma_y$ are Pauli's matrices. Having obtained eigenvalues and eigenfunctions, the order parameter can be calculated self-consistently from the gap equation,
\begin{eqnarray}
\Delta_{n}&=&\frac{g_0}{2}\sum_{k_{y(x)}}\sum_{\nu}(u_n^\nu(k_{y(x)})v_n^{\nu\ast}(k_{y(x)})\nonumber\\
&+& y_n^\nu(k_{y(x)})z_n^{\nu\ast}(k_{y(x)}))\tanh(\frac{E^\nu(k_{y(x)})}{2k_BT}),
\label{gap}
\end{eqnarray}
where $T$ is the temperature ($k_B$ is the Boltzmann constant).
Also, LDOS at zero temperature and thermodynamic potential~\cite{Thermo} can be calculated by the following equations,
\begin{eqnarray}
D_{n}^{ACG(ZCG)}(E)=\sum_{k_{y(x)}}\sum_{\nu}|\psi^\nu_n|^2\delta(E-E^\nu(k_{y(x)})), \label{LdFr1}\\
\Omega_S\!\!=\!-k_BT\!\!\sum_{k_{y(x)}}\sum_{\nu,\alpha=\pm}\!\!\!\ln[1\!+\!\exp({\frac{\alpha E^\nu(k_{y(x)})}{k_BT}})]\!\!+\!\!\sum_{n}\!\frac{\Delta^2_{n}}{g_0}.\label{LdFr}
\end{eqnarray}
In Eqs.(\ref{gap}), (\ref{LdFr1}) and (\ref{LdFr}), all positive eigenvalues are summed over. By integrating over the position of LDOS, one can determine DOS of system.

{\it Results and discussion.}---
With an initial guess of the order parameter, one can determine eigenvalues and eigenfunctions from Eq.(\ref{eq:diag}) and recalculate the order parameter from the gap equation [Eq.(\ref{gap})]. This process can be done iteratively till difference between successive values of the order parameters becomes smaller than a desired value. Throughout this paper, we set the hopping energy $t$ as the energy unit and the lattice constant a as the length unit. The s-wave attraction strength $g_0$ is fixed to be $1.36t$ and $\mu = 0.6t$ (unless otherwise specified). These values induce bulk gap 0.042t for the case of flat graphene sheet~\cite{Bogoliubov-Valatin1,Bogoliubov-Valatin2,Bogoliubov-Valatin3}. We are interested in the zero temperature case, where mean-field approach has the highest validity. Also, the number of unit cells for OBCs (PBCs) is chosen to be 199 (100), unless otherwise specified. In numerical calculation, we take the wavelength $\lambda = 30\textrm{a}$ which can be achievable experimentally~\cite{he}.
\begin{figure}[t]
  \centering
  \includegraphics[width=8.cm]{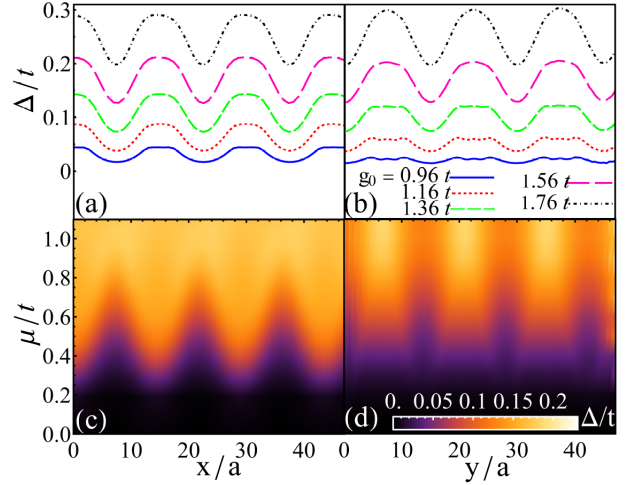}
  \caption{(Color online) Top panels (a) and (b) are spatial profiles of order parameter for $\mu = 0.6t$ and different values of $g_0$. Bottom panels (c) and (d) are density plot of order parameter versus position coordinates and $\mu$ for $g_0 = 1.36t$. Left (right) panels are for ACG (ZCG). Here $h_0 = 3\textrm{a}$ and open boundary conditions are applied over 400 unit cells.}
  \label{mug0}
\end{figure}

\begin{figure}[t]
  \centering
  \includegraphics[width=8.cm]{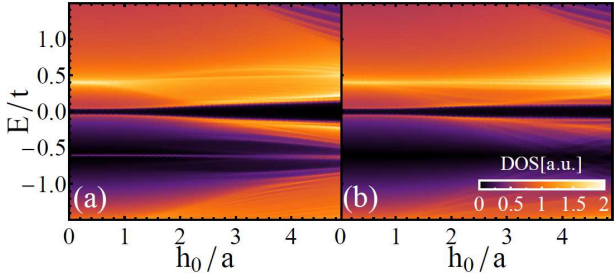}
  \caption{(Color online) Density plot of DOS in the superconducting state in terms of energy $E$ and ripple's amplitude $h_0$ for (a) ACG and (b) ZCG. The values of DOS are in the arbitrary unit. The dark region around the Fermi energy $E \approx 0$ indicates the energy gap. Here $g_0 = 1.36t$ and $\mu = 0.6t$.}
  \label{DOS}
\end{figure}

The spatial profile of order parameters in terms of $h_0$ and strain $u$ are shown in Figs.~\ref{cor}(c) and ~\ref{cor}(d) for the ACG and ZCG, respectively. With the weak to moderate out-of-plane deformation $h_0\leq2\textrm{a}$, induced superconductivity is rather homogenous, but, surprisingly, by further increasing $h_0$, the order parameter of induced s-wave superconductivity modulates spatially for both the ACG and ZCG. Also, there are two periods of the order parameter for both ACG and ZCG in the region between the two vertical dashed lines in Figs.~\ref{cor}(b), ~\ref{cor}(c) and ~\ref{cor}(d), which is equal to the ripple wavelength. As a consequence, the order parameter modulates with the half wavelength of the corrugation. However, there is a crucial difference. For the ZCG, the highest value of order parameter takes place at the dips and peaks of out-of-plane deformation, while the ACG has the lowest value of order parameter at such points. Although these results are calculated for $\lambda = 30 \textrm{a}$, we examined different wavelengths as well and  same qualitative results were found.

In Figs.~\ref{mug0} (a) and (b), the spatial profiles of order parameter of ACG and ZCG are shown for $h_0$ = 3a, $\mu$ = 0.6t and different values of $g_0$. We see that the amplitude and the magnitude of order parameter increase with increasing $g_0$. Panel c (d) of Fig.~\ref{mug0} refers to the dependence of order parameter of ACG (ZCG) on position coordinates x (y) and $\mu$ for $h_0$ = 3a and $g_0$ = 1.36t. As we increase $\mu$, the magnitude of order parameter increases to a certain value around van Hove singularity. In addition, the amplitude of order parameter modulation is vanishingly small around van Hove singularity in the case of ACG. Consequently, the effect of corrugation on superconducting modulation is significant over a considerable range of energies for the case of ZCG but in the case of ACG, the corrugation influences the superconducting state away from van Hove singularity. These two different behaviors arise from the band structure difference between ACG and ZCG. Here we have used 400 unit cells along open boundary conditions.

Using the self-consistent numerical values of order parameter, we evaluate the DOS of the superconducting state in the (E, $h_0$) plane, as shown in Fig.~\ref{DOS}. For both the ACG [see Fig.~\ref{DOS}(a)] and ZCG [see Fig.~\ref{DOS}(b)], superconducting gap around the Fermi surface increases with increasing $h_0$. It has been shown that Landau levels due to the pseudo magnetic field enhances superconductivity in strained graphene\cite{UniStrainUchoa}, but the condition $h_0^2/\lambda \textrm{a}\geq$1~\cite{1DLL} usually does not fulfill in the case of rippled graphene for generating 1D Landau levels, however increasing the amplitude of out-of-plane deformation enhances the order parameter up to one order of magnitude larger than that of flat graphene case even if $h_0^2/\lambda \textrm{a} < 1$. This can be attributed to the increase of degeneracy of states due to structural ripples as will be discussed in the following.
\begin{figure}[t]
  \centering
  \includegraphics[width=8.cm]{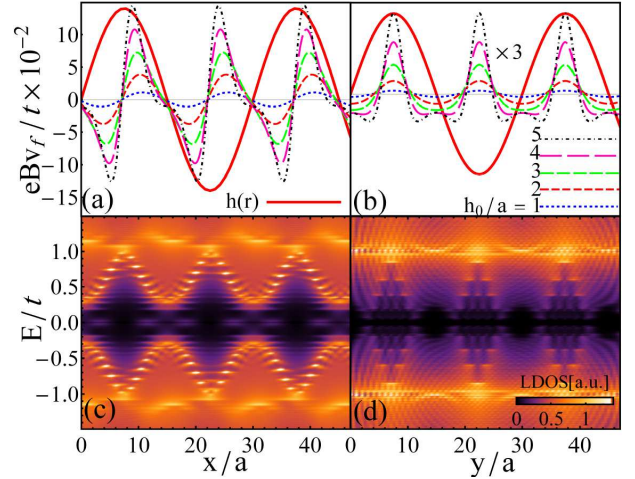}
  \caption{(Color online) Top panels (a) and (b) are spatial profiles of pseudo magnetic field for different values of $h_0$. Bottom panels (c) and (d) are density plots of LDOS for $h_0 = 4.5\textrm{a}$. Left (right) panels are for ACG (ZCG). For illustrative purposes the profile of out-of-plane deformation h(r) is shown in panels (a) and (b) with solid red line. Here $g_0 = \mu = 0t$.}
  \label{PseLDOS}
\end{figure}

In order to describe these behaviors, we focus on investigating the effect of non-uniform pseudo magnetic field generated by 1D periodic ripples on the electronic properties of normal state for pure graphene. Lattice deformations modify the hopping energies and consequently modify the band structure. One can translate this modified band structure into unmodified band structure plus strain-induced pseudo vector potential~\cite{graphene2}. The lowest order of pseudo vector potential around Dirac point $\mathbf{K} = (0,-1)4\pi/(3\textrm{a})$ can be approximated by~\cite{PseudoVectorAppro},
\begin{eqnarray}
\mathbf{A}_{i}=\frac{1}{2ev_f}\left(
                                  \begin{array}{c}
                                    \sqrt{3}(t_{i 2}-t_{i 3}) \\
                                    t_{i 2}+t_{i 3}-2t_{i 1}  \\
                                  \end{array}
                                \right),
\label{pesu}
\end{eqnarray}
where $v_f$ is the Fermi velocity and $e$ is the electron charge. The pseudo magnetic field can be calculated by $\mathbf{B}=\mathbf{\nabla}\times \mathbf{A}_{i}$.
In Figs.~\ref{PseLDOS}(a) and ~\ref{PseLDOS}(b), plots of the pseudo magnetic field are shown for the ACG and ZCG, respectively, with different values of $h_0$ and for illustrative purposes the profile of out-of-plane deformation $h(r)$ is shown as well. Also, LDOSs of the ACG and ZCG are depicted in Figs.~\ref{PseLDOS}(c) and ~\ref{PseLDOS}(d) for $h_0 = 4.5\textrm{a}$.
In the case of ACG, the pseudo magnetic field changes sign abruptly at the peak and dip of ripples (see Fig.~\ref{PseLDOS}(a)) and the corresponding points in the LDOS plot have no states around the Fermi surface, while at the middle points between the peaks and dips of the out-of-plane deformation, the change of pseudo magnetic field is smooth and the corresponding points in the LDOS have finite states as shown in Fig.~\ref{PseLDOS}(c). The situation is reversed in the ZCG, so that the pseudo magnetic field (see Fig.~\ref{PseLDOS}(b)) and LDOS around the Fermi surface (see Fig.~\ref{PseLDOS}(d)) have considerable values at the peaks and dips of ripples. These behaviors can be related to the manner of stretching of bounds' length and correspondingly the changes of $t_{i\alpha}$'s. In the case of ACG, $t_{i2}$ and $t_{i3}$ change identically but the change of $t_{i1}$ is more dominant due to corrugation. As the amplitude of out-of-plane deformation increases, the changes of $t_{i2}$ and $t_{i3}$ become effective resulting in asymmetric pseudo magnetic field. For the case of ZCG, the corrugation only changes $t_{i2}$ and $t_{i3}$ in a same way with a phase difference. Thus the resulting pseudo magnetic field becomes symmetric. Consequently, carriers in the corrugated graphene are affected by non-uniform pseudo magnetic field.

\begin{figure}[t]
  \centering
  \includegraphics[width=8.cm]{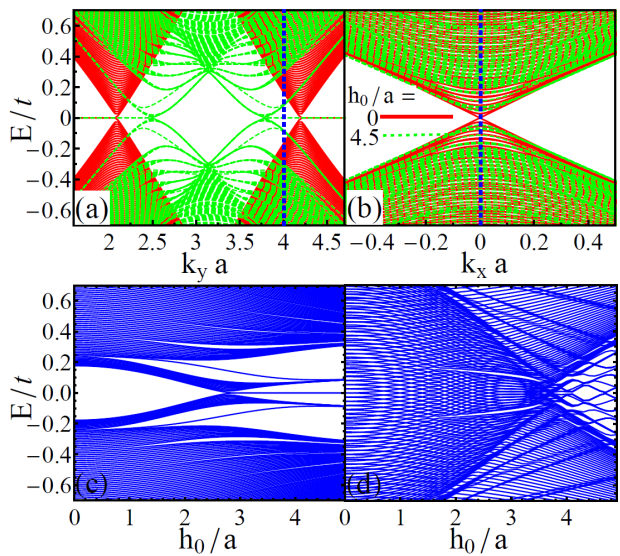}
  \caption{(Color online) Top panels (a) and (b) are band structures for two values $h_0 = 0\textrm{a}$ (solid red lines) and $h_0 = 4.5\textrm{a}$ (dashed green lines). Bottom panels (c) and (d) are the evolution of energy levels versus $h_0$ at the crystal momentum $k_y = 4/\textrm{a}$ and $k_x = 0$ indicated by vertical dashed lines in the top panels. Left (right) panels are for the ACG (ZCG).}
  \label{Band}
\end{figure}

The presence of the non-uniform pseudo magnetic field strongly influence the energy states of the system. The band structures of the ACG and ZCG are presented in Figs. ~\ref{Band}(a) and ~\ref{Band}(b) for two values $h_0 = 0\textrm{a}$ and $h_0 = 4.5\textrm{a}$. For the ACG, Dirac points shift towards each other while for the ZCG case, the location of Dirac point remains unchanged. In both cases, the energy levels at some states become degenerate for $h_0 = 4.5\textrm{a}$. The evolution of energy levels versus $h_0$ at the crystal momentum $k_y=4/\textrm{a}$ and $k_x = 0$, as indicated by vertical dashed lines in Figs.~\ref{Band}(a) and ~\ref{Band}(b), is shown in Figs.~\ref{Band}(c) and ~\ref{Band}(d) for ACG and ZCG, respectively. 
We see that as $h_0$ increases, the energy states join together and become degenerate, depending on the values of $E$ and $k_{y(x)}$.

\begin{figure}[t]
  \centering
  \includegraphics[width=5.cm]{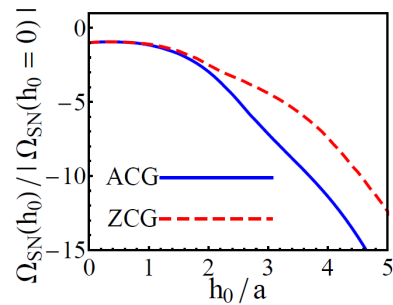}
  \caption{(Color online) The difference of thermodynamic potentials of superconducting and normal states $\Omega_{SN} = \Omega_S-\Omega_N$ versus $h_0$ normalized with absolute value of $\Omega_{SN}$ of flat graphene case $h_0$ = 0.}
  \label{Free}
\end{figure}

To check the stability of IS emerging from periodic ripples, we evaluate the thermodynamic potential of superconducting state $\Omega_S$ (Eq.(\ref{LdFr})) with the order parameters determined self-consistently. Similarly, the thermodynamic potential of normal state $\Omega_N$ can be determined by setting $\Delta = 0$. The difference of the thermodynamic potentials of the superconducting and normal states $\Omega_{SN} = \Omega_S-\Omega_N$ versus $h_0$ is illustrated in Fig.~\ref{Free}. As $h_0$ increases, the thermodynamic potential of superconducting state becomes smaller than the normal state for both the ACG and ZCG indicating that ripples drive uniform superconducting state towards a more stable IS. Also, in the range $h_0 < 2\textrm{a}$, the $\Omega_{SN}$'s of both ACG and ZCG are the same, but at larger values of $h_0$, the $\Omega_{SN}$ of ACG is smaller than the ZCG.

In order to establish superconductivity in a corrugated graphene layer, we propose to deposit graphene layer on top of a superconducting substrate~\cite{graphOnTop} decorated with parallel horizontally aligned superconducting cylinders for producing out-of-plane deformation. It is well-known that graphene layer has intrinsic ripples which do not allow the graphene' surface to follow the substrate corrugations~\cite{NeekCorr1,NeekCorr2}, therefore superconductivity can be locally induced in the graphene by proximity effect through the regions of the graphene layer that are in contact with the superconductor. However, it has been shown that if the distance between local superconductivity is smaller than superconducting coherence length, even local superconductivity can establish superconductivity throughout the graphene layer~\cite{SuperDecor1,SuperDecor2,SuperDecor3}. The van der Waals interaction sticks the graphene sheet to both the superconducting substrate and islands causing a non-uniform strain. Also, recent advances in fabrication of graphene membrane have made a situation that the wavelength, the amplitude and the orientation of out-of-plane deformation of a graphene membrane can be controllably produced \cite{ripple1,ripple2,ripple3,ripple4,ripple5}.
\par
{\it Conclusions.}---We studied new type of inhomogeneous superconducting state in the presence of time-reversal symmetry. The interplay of non-uniform pseudo magnetic field, generated by honeycomb lattice deformation, and superconducting correlations results in IS. We employed 1D periodic array of structural ripples along the armchair or zigzag directions as the source of non-uniform pseudo magnetic field in a graphene layer and found that order parameter of superconducting state with s-wave symmetry modulates spatially. The manner of modulation strongly depends on the corrugation direction. Also, the increase of amplitude of out-of-plane deformation leads the system towards a more stable superconducting state.

\acknowledgments
The author is grateful to A. Ghorbanzadeh and A. Najafi for fruitful discussions and comments.

\end{document}